\documentclass[12pt]{article}
\usepackage{graphicx}
\usepackage{amssymb,epsf}
\csname @addtoreset\endcsname{equation}{section}

\def\bseq{\begin{subequation}}  
\def\eseq{\end{subequation}}
\def\bsea{\begin{subeqnarray}}  
\def\esea{\end{subeqnarray}}

\newcommand{\bbox}{\lower.2ex\hbox{$\Box$}}

\newcommand{\beq}{\begin{equation}}
\newcommand{\eeq}{\end{equation}}
\newcommand{\bea}{\begin{eqnarray}}
\newcommand{\eea}{\end{eqnarray}}
\newcommand{\ena}{\end{eqnarray}}

\renewcommand{\a}{\alpha}
\renewcommand{\b}{\beta}

\renewcommand{\d}{\delta}

\newcommand{\g}{\gamma}
\newcommand{\G}{\Gamma}

\newcommand{\e}{\epsilon}
\newcommand{\z}{\zeta}

\newcommand{\p}{\pi}

\newcommand{\Db}{\bar{D}}

\newcommand{\Phib}{\bar{\Phi}}

\begin{document}
\begin{titlepage}
\begin{flushright}
IFUM-866-FT \\
Bicocca-FT-06-10 \\
\end{flushright}
\vspace{.3cm}
\begin{center}
{\Large \bf  Conformal invariance of the  planar\\

\vspace{0.2cm}
 $\b$--deformed
${\cal N}=4$ SYM theory requires $\b$ real}
\vfill

{\large \bf Federico Elmetti$^1$,~Andrea Mauri$^1$,~Silvia
Penati$^2$, \\

\vspace{0.2cm} Alberto Santambrogio$^1$ and
Daniela Zanon$^1$}\\

\vspace{0.5cm}

{\small $^1$ Dipartimento di Fisica, Universit\`a di Milano and\\
INFN, Sezione di Milano, Via Celoria 16, I-20133 Milano, Italy\\

\vspace{0.1cm}
$^2$ Dipartimento di Fisica, Universit\`a di Milano--Bicocca and\\
INFN, Sezione di Milano--Bicocca, Piazza della
Scienza 3, I-20126 Milano, Italy}\\
\end{center}
\vfill
\begin{center}
{\bf Abstract}
\end{center}
{
  We study the  ${\cal N}=1$ $SU(N)$ SYM theory which is a
marginal deformation of the ${\cal N}=4$ theory, with a complex
deformation parameter $\b$. We consider the large $N$ limit and
study perturbatively  the conformal invariance condition. We find
that finiteness requires reality of the deformation parameter
$\b$. } \vspace{2mm} \vfill \hrule width 3.cm
\begin{flushleft}
e-mail: federico.elmetti@mi.infn.it\\
e-mail: andrea.mauri@mi.infn.it\\
e-mail: silvia.penati@mib.infn.it\\
e-mail: alberto.santambrogio@mi.infn.it\\
e-mail: daniela.zanon@mi.infn.it
\end{flushleft}
\end{titlepage}

The ${\cal N}=4$ supersymmetric Yang-Mills theory offers one of
the best playgrounds to test new ideas connected to nonperturbative 
and exact results. Using the AdS/CFT correspondence \cite{M}
it has allowed to get new insights and a deeper understanding of
duality properties enjoyed by the gauge theory and the
corresponding supergravity. The search for theories with a less
degree of supersymmetry that nonetheless might possess features
similar to the ones of  ${\cal N}=4$ SYM has lead to consider
theories obtained deforming the ${\cal N}=4$ theory itself. Of
special interest are ${\cal N}=4$ marginal deformations analyzed in
\cite{LS} for which the supergravity dual description has been
found in \cite{LM}.

In this paper we consider such marginal deformations. They are called 
$\beta$--deformations since they are obtained 
by modifying the original ${\cal N}=4$ superpotential for the
chiral superfields in the following way 
\beq 
ig~{\rm Tr}(  ~\Phi_1
\Phi_2 \Phi_3 - ~ \Phi_1 \Phi_3 \Phi_2 ~)~\longrightarrow ~ih~{\rm
Tr}\left( ~e^{i\pi\b} ~\Phi_1 \Phi_2 \Phi_3 - e^{-i\pi\b}~ \Phi_1
\Phi_3 \Phi_2 ~\right) 
\label{deformation} 
\eeq 
where in general
$h$ and $\b$ are complex constants. In \cite{LS} it was argued
that these $\b$-deformed ${\cal N}=1$ theories become conformally
invariant, i.e. the deformation becomes exactly marginal, if one
condition is satisfied by the constants $h$ and $\b$. For the case
of $\b$ real and in the planar limit it has been shown \cite{MPSZ}
that the condition 
\beq h\bar{h}=g^2 
\label{betareal} 
\eeq 
ensures
conformal invariance of the theory to all perturbative orders and provides
the exact field theory dual to the Lunin--Maldacena supergravity 
background \cite{LM}. 

The aim of the present investigation is to study how the conformal
invariance condition can be implemented for the case of complex
$\b$. The analysis is done using a perturbative approach and
imposing the finiteness of the two-point chiral correlators. In
turn this guarantees the vanishing  of all the $\b$-functions \cite{beta}. 
We find that in the planar limit conformal invariance is achieved
only for {\em real} values of the parameter $\b$. 
This result seems to be in direct correspondence with the findings of the
string dual approach in which singular solutions are produced 
whenever $\b$ acquires a non vanishing imaginary part \cite{LM,FRT,GN,KT}. 
We will comment on this in our conclusions.

\vspace{0.8cm} 
In order to perform higher order perturbative
calculations it is very efficient to rely on ${\cal N}=1$
superspace techniques. In this setting the $\b$--deformed theory
is described by the following action (we  use notations and
conventions as in \cite{superspace}, see also \cite{PSZ2}) 
\bea 
S &=&\int d^8z~ {\rm
Tr}\left(e^{-gV} \Phib_i e^{gV} \Phi^i\right)+ \frac{1}{2g^2}
\int d^6z~ {\rm Tr} (W^\a W_\a)\nonumber\\
&&+ih  \int d^6z~ {\rm Tr}( ~q ~\Phi_1 \Phi_2 \Phi_3 - q^{-1}~
\Phi_1 \Phi_3 \Phi_2 ~)
\nonumber \\
&& + i\bar{h}\int d^6\bar{z}~ {\rm Tr} ( ~{\bar{q}}^{-1}~ \Phib_1
\Phib_2 \Phib_3 - \bar{q} ~\Phib_1 \Phib_3 \Phib_2~ )\qquad,\qquad
q\equiv e^{i\pi\b}  \label{actionYM} 
\eea 
where $h$ and $\b$ are
complex couplings and $g$ is the real gauge coupling constant. The
superfield strength $W_\a= i\Db^2(e^{-gV}D_\a e^{gV})$ is given in
terms of a real prepotential $V$, while $\Phi_i$ with $i=1,2,3$
are the three chiral superfields of the original ${\cal N}=4$ SYM
theory.  We write $V=V^aT_a$, $\Phi_i=\Phi_i^a T_a$ where $T_a$
are $SU(N)$ matrices in the fundamental representation. In the
undeformed theory one has $h=g$ and $q=1$.

We want to study the condition that the couplings have to satisfy
in order to guarantee the conformal invariance of the theory for
complex values of $h$ and $\b$ in the large $N$ limit. As observed
above to this end it is sufficient to impose the finiteness on the two-point 
chiral correlator \cite{beta}.

In the large $N$ limit for real values of $\b$, i.e. if the
condition $q\bar{q}=1$ is satisfied,  the $\b$-deformed theory
becomes exactly conformally invariant if the condition (\ref{betareal})
is satisfied \cite{MPSZ}. 
This means that if the chiral couplings differ only
by a phase from the ones of the ${\cal N}=4$ SYM theory, the
planar limits of the two theories are essentially the same (see
also \cite{Khoze}).

When $q\bar{q}\neq 1$ the easiest way \cite{RSS} to study the condition of
conformal invariance is to  look at the difference between the
two-point $\b$-deformed correlator and the corresponding one in
the ${\cal N}=4$ SYM theory. If we want to have an
exactly marginal deformation the difference must be finite. We
will proceed perturbatively in superspace. The propagators for the
vector and chiral superfields, and the interaction vertices are
obtained directly from the action in (\ref{actionYM}). Supergraphs
are evaluated performing standard $D$-algebra in the loops and the
corresponding divergent integrals are computed using dimensional
regularization in $D=4-2\e$.

\begin{figure}[ht]
\begin{center}
\epsfysize=2.8cm\epsfbox{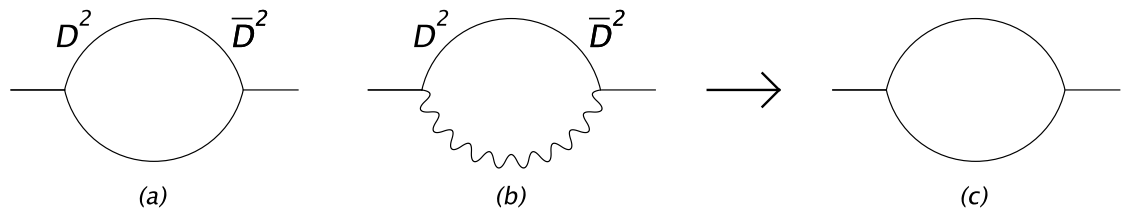}
\end{center}
\caption{Supergraphs contributing at one loop}
\label{Fig1}
\end{figure} 

At one loop the analysis is very simple and mimics exactly what
happens in the $\b$ real case \cite{FG, PSZ1, RSS, MPSZ}. 
The divergent supergraphs are shown in Fig.1. 
The chiral field propagators are given by 
\beq 
\langle \Phi_i\Phib_j \rangle=
-\d_{ij}\frac{1}{\Box}=\d_{ij}\frac{1}{p^2} 
\label{propchiral}
\eeq 
while the vector propagators are 
\beq 
\langle VV \rangle =\frac{1}{\Box}=-\frac{1}{p^2} 
\label{propvector} 
\eeq 
The D-algebra is the same for the two configurations and its
completion gives rise to a logarithmically divergent momentum
integral. The diagrams (with different color configurations) in 
Fig.1b containing a vector line are the
same in the ${\cal N}=4$  and in the $\b$-deformed SYM theory,
since they only depend on the gauge coupling $g$. The diagrams in
Fig.1a  contain the chiral couplings:  in the deformed theory they
give a contribution \beq
\frac{N}{(4\p)^2}~h\bar{h}~\left( q\bar{q}+\frac{1}{q\bar{q}}\right) ~
\frac{1}{\e} \label{1loopdef} \eeq
 while  in the
${\cal N}=4$ theory  they are proportional to $g^2$ \beq
\frac{N}{(4\p)^2}~2 g^2 ~\frac{1}{\e} \label{1loopN4} \eeq In
order  to achieve finiteness one has to impose that the difference
between the two results be finite. This implies that to this order
the $\b$-deformed theory is conformal invariant if \beq
h\bar{h}~\left( q\bar{q}+\frac{1}{q\bar{q}}\right) =2 g^2 \label{finite} \eeq

Now we consider higher-loop contributions. Since we look at the 
difference between the two-point correlators computed in the 
$\b$-deformed theory and in the ${\cal N}=4$ SYM, we need not consider 
diagrams that contain only gauge-type
vertices their contributions being the same in the two
theories.  Therefore we concentrate on divergent graphs that
contain either only chiral vertices or mixed chiral and gauge
vertices. Moreover we observe that a chiral loop can close only if
it has the same number of chiral and antichiral vertices, i.e. no
polygonal configurations with an odd number of vertices are
possible. With these rules in mind it is straightforward to
analyze the two- and three-loop contributions. At two loops we
have the diagrams shown in Fig.2.  

\begin{figure}[ht]
\begin{center}
\epsfysize=6cm\epsfbox{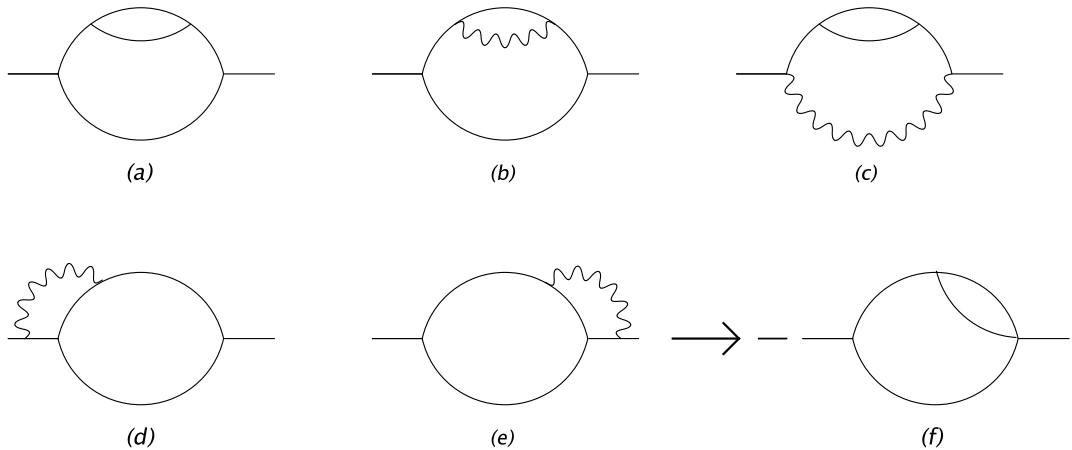}
\end{center}
\caption{Supergraphs contributing at two loops}
\label{Fig2}
\end{figure} 

\noindent
For all the
different configurations the D-algebra leads to the same bosonic integral
 in Fig.2f.
It is very simple to compute the various color factors: we have
for the $\b$-deformed theory 
\bea
Fig.2a&\longrightarrow& -2\left[h\bar{h}\left(q\bar{q}+\frac{1}{q\bar{q}}
\right)\right]^2 N^2\nonumber\\
&& \nonumber\\
Fig.2b+2c+2d+2e&\longrightarrow&
2\left[h\bar{h}\left(q\bar{q}+\frac{1}{q\bar{q}}\right)\right]g^2
N^2 \label{2Lgraphbeta} \eea while correspondingly for ${\cal
N}=4$ SYM we find \bea
Fig.2a&\longrightarrow& -8g^4  N^2\nonumber\\
Fig.2b+2c+2d+2e&\longrightarrow& 4 g^4 N^2 \label{2LgraphN4} \eea
If we compute the difference of the results in (\ref{2Lgraphbeta})
and in (\ref{2LgraphN4}) and use the conformal invariance
condition in (\ref{finite}), we obtain a zero result. This means
that the condition we found at one loop ensures finiteness also at
two loops. In fact repeating a similar analysis  at three loops
one can easily show that (\ref{finite})  makes the divergent
diagrams computed in the deformed theory equal to the
corresponding ones in the ${\cal N}=4$ SYM.  In the planar limit
under the condition in (\ref{finite}) the two-point correlators do
coincide up to three loops. Up to this order the situation is
completely parallel to the case of the real $\b$-deformation \cite{PSZ1, RSS}:
there $q\bar{q}=1$ and the condition in (\ref{finite}) was simply
given by $h\bar{h}=g^2$. This condition
was actually sufficient \cite{MPSZ} to implement finiteness of the two-point
correlator in the planar limit
to {\it all orders} in perturbation theory . Moreover  the two-point correlator of the real
$\b$-deformed theory becomes {\it exactly} equal to the one
computed in the ${\cal N}=4$ theory.

Now we proceed in the study of the $\b$-complex case and examine
the situation at four loops. We will find that at this order we
are forced to modify the condition in (\ref{finite}). This should
not come as a surprise because of the following reason: as
explained above the divergence at one loop is linked to the color
factor of the chiral bubble in Fig.1a and this leads to the
condition in (\ref{finite}).  At two and three loops divergent
graphs are constructed either by inserting vector lines on chiral
bubbles or by assembling chiral bubbles together. Since the
addition of vectors simply modifies the color due to the chiral
vertices by the multiplication of $g^2$ factors, in both cases the
condition in (\ref{finite}) suffices to give conformal invariance.
In fact this same reasoning applies also to all the four-loop 
diagrams that either contain vector lines on chiral bubbles or
consist of various arrangements of chiral bubbles: for all these cases
the condition in (\ref{finite}) makes these graphs equal to the corresponding 
ones in the ${\cal N}=4$ theory. The novelty is that
at four loops a new type of chiral divergent structure does arise.
We  will be able to implement the cancelation of divergences at
order $g^8$, but in contradistinction to the real $\b$ case
finite parts will survive in the $\b$-deformed two-point function
which are absent in the corresponding ${\cal N}=4$ two-point
function.  

\begin{figure}[ht]
\begin{center}
\epsfysize=2.8cm\epsfbox{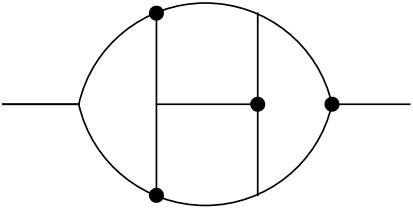}
\end{center}
\caption{New planar structure at four loops; the vertices with dots are 
antichiral}
\label{Fig3}
\end{figure} 

\begin{figure}[ht]
\begin{center}
\epsfysize=4.5cm\epsfbox{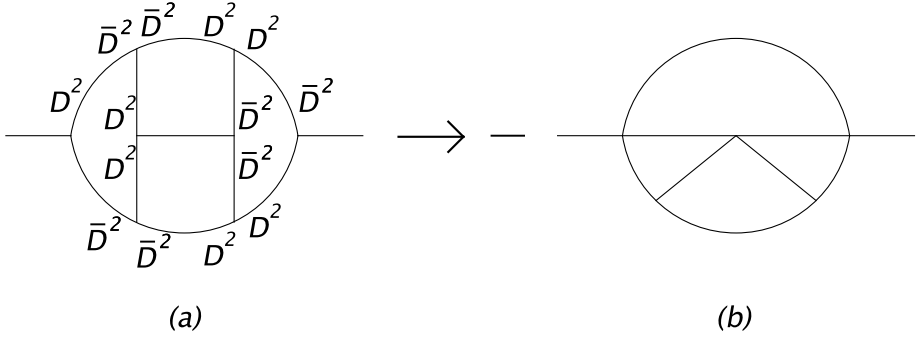}
\end{center}
\caption{D-algebra for the supergraph in Fig.3}
\label{Fig4}
\end{figure} 

The new type of chiral supergraph, i.e. not containing chiral
bubble insertions, is the one drawn in Fig.3. The D-algebra
structure shown explicitly  in Fig.4a is the same for all the
arrangements of the three chiral superfields at the vertices.
Completing the D-algebra in the loops one obtains the  bosonic
graph shown in Fig.4b. The corresponding integral is divergent
\cite{Kaz}
\bea 
I_4&=&-\int \frac{d^D k ~d^D q ~d^D r ~d^D t
}{(2\p)^{4D}}\frac{1}{k^2 (k+t)^2 (q+k)^2 (q+r)^2 (q+p)^2
t^2 r^2 (t+r)^2}\nonumber\\
&&~~~\nonumber\\
&=& ~-5~\z(5)~\frac{1}{(4\p)^8}\frac{1}{\e}~\frac{1}{(p^2)^{4\e}}
\label{4loop} 
\eea 
The color factor is also easily computed: one
has to sum over all the various possibilities at the chiral
vertices and in so doing one finds 
\beq
C_4=~N^4~(h\bar{h})^4\left[(q\bar{q})^4+\frac{1}{(q\bar{q})^4}+6\right]
\label{color4} 
\eeq 
The factor in (\ref{color4}) can be rewritten as 
\beq
C_4=~\frac{N^4}{2}~(h\bar{h})^4\left[\left(q\bar{q}+\frac{1}{q\bar{q}}\right)^4
+ \left(q\bar{q}-\frac{1}{q\bar{q}}\right)^4\right] 
\label{4color}
\eeq 
In this way it is easy to compare the result with the one we
would have obtained in ${\cal N}=4$ SYM. In fact using the
condition in (\ref{finite}) we find that the $\b$-deformed
two-point function at four loops differs from the corresponding
${\cal N}=4$ two-point function by the contribution
\beq
J_4=~-\frac{5}{2}~\z(5)~N^4~\frac{1}{(4\p)^8}~
\frac{1}{\e}~\frac{1}{(p^2)^{4\e}}
(h\bar{h})^4\left(q\bar{q}-\frac{1}{q\bar{q}}\right)^4
\label{4diff} 
\eeq 
If we want the $\b$-deformed theory to be
conformally invariant this term has to be cancelled. The only way
out is to modify the relation of conformal invariance in
(\ref{finite}), so that a contribution from a lower-loop order
might cancel the unwanted four-loop divergence.

In the spirit of \cite{LS} (see also \cite{previous}), in the space 
of the coupling constants we are looking for a surface of
renormalization group fixed points. To this end we set 
\beq
h_1\equiv h q\qquad\qquad h_2\equiv \frac{h}{q} 
\eeq 
and reparametrize these couplings in terms of the gauge coupling $g$.
In fact since in the planar limit for each diagram the color factors 
from chiral vertices is always in terms of  the products $h_1^2\equiv
h_1\bar{h}_1$ and $h_2^2\equiv h_2\bar{h}_2$ we express directly
$h_1^2$ and  $h_2^2$ as power series in the coupling $g^2$ as
follows 
\bea
&& h_1^2=a_1 g^2+a_2 g^4+a_3 g^6+\dots \nonumber\\
&&h_2^2=b_1 g^2+b_2 g^4+b_3 g^6+\dots 
\label{expansion} 
\eea
The coefficients  $a_i$ and $b_i$ will be determined by imposing
that what we obtain from various loop orders, subtracted by the
corresponding ${\cal N}=4$ results, vanishes order by order in the  
$g^2$ expansion.

In order to make the comparison with the ${\cal N}=4$ calculation
simpler we find convenient to determine the general structure of the 
color factors of the relevant diagrams. At L--loop order the color factor
is a homogeneous polynomial in $h_1^2, h_2^2$ and $g^2$ of degree $L$. 
Moreover, as a consequence of the invariance of the theory under the global 
symmetry $h_1 \leftrightarrow -h_2$ and 
$\Phi_i \leftrightarrow \Phi_j$, $i \neq j$, it has to be symmetric under
$h_1^2 \leftrightarrow h_2^2$. These properties, together with the requirement
of having a smooth limit to $(2g^2)^L$ in the ${\cal N}=4$ limit 
($h_1^2, h_2^2 \rightarrow g^2$), constrain the $L$--loop color factor 
to have the following form  
\footnote{We do not worry about an overall normalization factor since it is 
irrelevant for our general argument}
\beq
F^{(L)}(h_1^2+h_2^2) + (h_1^2-h_2^2)^2~ G^{(L-2)}(h_1^2,h_2^2) 
\label{natural} 
\eeq
with $F^{(L)}(2g^2) = (2g^2)^L$. The functions $F^{(L)}$ and $G^{(L-2)}$ 
depend also on the coupling $g^2$, but for notational simplicity 
we have  chosen not to write it explicitly. They are homogeneous polynomials 
of degrees $L$ and $(L-2)$ respectively, symmetric in $h_1^2, h_2^2$. 
Their general form is 
\bea
&& F^{(L)}(h_1^2+h_2^2) = \sum_{k=0}^{L}  (h_1^2+h_2^2)^k ~ 
(2g^2)^{L-k}~f_k
\nonumber \\
&& G^{(L-2)}(h_1^2,h_2^2) = \sum_{k=0}^{[(L-2)/2]} (h_1^2-h_2^2)^{2k} ~ 
{\cal P}^{(L-2-2k)}(h_1^2,h_2^2)
\label{functions}
\eea
with constant coefficients $f_k$ satisfying $\sum_{k=0}^{L} f_k =1$  
and ${\cal P}^{(L-2-2k)}$ homogeneous polynomials not vanishing 
for $h_1^2 = h_2^2$.

We note that for pure chiral diagrams, the ones we will be mainly interested
in, there is no $g^2$--dependence in $F^{(L)}$ and $G^{(L-2)}$ and, 
in particular, $F^{(L)}(h_1^2+h_2^2) = (h_1^2+h_2^2)^L$. 

At $L$--loop order, after we take the difference with the ${\cal N}=4$ result
what is left over is given by 
\beq
\G^{(L)}=\left[F^{(L)}(h_1^2+h_2^2) - (2g^2)^L +
(h_1^2-h_2^2)^2 ~G^{(L-2)}(h_1^2,h_2^2)\right]~I_{div}^{(L)}
\label{leftover} 
\eeq 
where $I_{div}^{(L)}$ denotes the divergent
factor from the L-loop integral. 
Finally summing over all loops and
using the expansions in (\ref{expansion}) we end up with 
\bea
\sum_L ~\G^{(L)}&=&\sum_L ~\left[F^{(L)}(h_1^2+h_2^2)-(2g^2)^L+
(h_1^2-h_2^2)^2~G^{(L-2)}(h_1^2,h_2^2)\right]~I_{div}^{(L)}\nonumber\\
&=&\sum_k  A_k ~(g^2)^k 
\label{finalexp} 
\eea 
Conformal invariance is achieved imposing 
\beq 
A_k=0 
\label{confinv} 
\eeq 
order by order in $g^2$.

Thus we go back to the one-loop calculation and apply concretely
the general procedure described above. From the results quoted in
(\ref{1loopdef}) and (\ref{1loopN4}) we see that $G^{(-1)}=0$ and
find \beq
\G^{(1)}=\left[F^{(1)}(h_1^2+h_2^2)-(2g^2)\right]~I_{div}^{(1)}=
\frac{N}{(4\p)^2}~\left[
h_1^2+h_2^2-2g^2\right]~ \frac{1}{\e} \label{leftover1loop} \eeq
 Therefore using the expansions in
(\ref{expansion}) at order $g^2$ we have to impose the condition
\beq {\cal O}(g^2): \qquad \qquad A_1=0 \qquad \longrightarrow
\qquad a_1+b_1-2=0 \label{order1} \eeq In fact since we have shown
that the condition in (\ref {finite}) ensures conformal invariance
up to three loops, up to order $g^6$,
 we find the following additional requirements 
\bea
&& {\cal O}(g^4): \qquad \qquad A_2=0 \qquad \longrightarrow \qquad a_2+b_2=0 
\nonumber\\
&&{\cal O}(g^6): \qquad \qquad A_3=0 \qquad \longrightarrow \qquad
a_3+b_3=0 \label{order2-3} 
\eea 
At this point it should be clear
that, according to the procedure we have illustrated  above, we do
not need consider anymore diagrams containing insertions of chiral bubbles
like the one in Fig.1a: once the condition (\ref{order1}) is
satisfied these diagrams do not lead to new divergent contributions.
Therefore at every loop order we have to isolate diagrams
corresponding to new chiral structures with eventually vector
propagators inserted on them.

Now we reexamine the results we have obtained up to four loops,
i.e.  up to order $g^8$. From the four-loop calculation (see eqs.
(\ref{4loop}) and (\ref{4color})) we have
\bea 
&& -\frac{5}{2}~\z(5)~N^4~\frac{1}{(4\p)^8}~\frac{1}{\e}~
(h\bar{h})^4\left\{ \left(q\bar{q}+\frac{1}{q\bar{q}}\right)^4
+\left(q\bar{q}-\frac{1}{q\bar{q}}\right)^4\right\}\nonumber\\
&=&
-\frac{5}{2}~\z(5)~N^4~\frac{1}{(4\p)^8}~\frac{1}{\e}~\left[(h_1^2+h_2^2)^4+
(h_1^2-h_2^2)^4\right] 
\label{4loop2} 
\eea 
Therefore we find \beq
\G^{(4)}=-\frac{5}{2}~\z(5)~N^4~\frac{1}{(4\p)^8}~\frac{1}{\e}~
\left[(h_1^2+h_2^2)^4-(2g^2)^4+
(h_1^2-h_2^2)^4\right] 
\label{leftover4loop} 
\eeq 
Now we insert
into (\ref{finalexp}) the results we have found so far, i.e.
(\ref{leftover1loop}) and (\ref{leftover4loop}) and use the
expansions in (\ref{expansion}) with the conditions in
(\ref{order1}) and (\ref{order2-3}). In this way  we find that the
conformal invariance condition at order $g^8$ is satisfied if \beq
{\cal O}(g^8): \qquad \quad A_4=0 \quad \longrightarrow \quad
a_4+b_4-\frac{5}{2}~\z(5)~N^3~\frac{1}{(4\p)^6}(a_1-b_1)^4=0
\label{g8} \eeq Up to this point we have ensured that the
two-point function is finite up to the order $g^8$. The finite
contributions explicitly depend on $q$ and vanish in the
corresponding terms of the ${\cal N}=4$ theory.

\begin{figure}[ht]
\begin{center}
\epsfysize=5.6cm\epsfbox{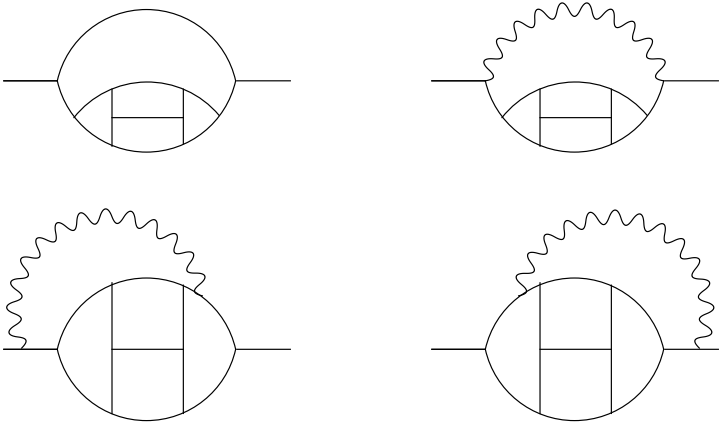}
\end{center}
\caption{Planar supergraphs with $1/\e^2$ divergences at five loops}
\label{Fig5}
\end{figure} 

The next step leads us to order $g^{10}$: we have to consider the
new five-loop diagrams and the two-loop diagrams  that will talk
to the five-loop graphs once the conformal invariance condition
(\ref{g8}) is imposed. Following the procedure described so far,
i.e. implementing the conformal invariance condition order by
order in the couplings, at the order $g^8$ we ended up adding
contributions coming from one-loop integrals and from four-loop
integrals. Now these structures show up at order $g^{10}$ as
subdivergences in two-loop and five-loop integrals respectively
and  they are responsible for the insurgence of $1/\e^2$-pole
terms. In fig.2 and in Fig.5 we have drawn  the two- and five-loop
diagrams which give rise to $1/\e^2$-pole terms. Having cancelled
divergences at lower orders one might be tempted to believe that
these $1/\e^2$ terms would automatically add up to zero. Indeed
this would be the case if we were cancelling divergences order by
order in loops. As emphasized above we are proceeding order by
order in the coupling $g^2$. At the order $g^8$ imposing the
relation (\ref{g8}) we have cancelled the $1/\e$ pole from the
one-loop diagram in Fig.1c with the $1/\e$ pole appearing from the
graph at four loops in Fig.4b. Essentially if we write
schematically the one-loop result as \beq A~\frac{1}{\e}~\frac
{1}{(p^2)^\e} \label{one} \eeq and the four-loop result as \beq
B~\frac{1}{\e}~\frac {1}{(p^2)^{4\e}} \label{four} \eeq imposing
the relation in (\ref{g8}) we have set $A+B=0$. When we go one loop
higher we have to deal with the bosonic integrals shown in Fig.6.

\begin{figure}[ht]
\begin{center}
\epsfysize=4.8cm\epsfbox{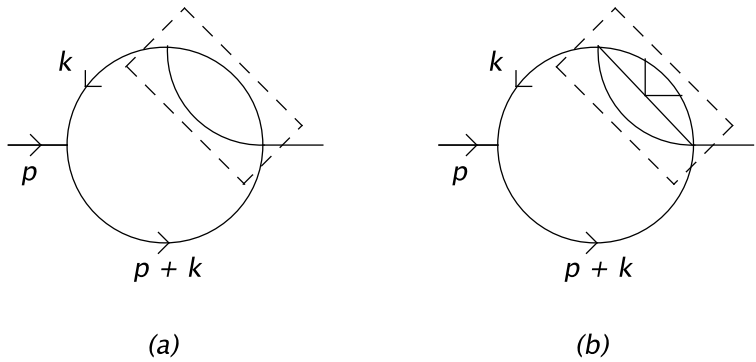}
\end{center}
\caption{Subtraction of subdivergences at order $g^{10}$}
\label{Fig6}
\end{figure} 

The $1/\e^2$ term in Fig.6a arises from \beq A~\frac{1}{\e}\int
d^D k \frac{1}{(p+k)^2 (k^2)^{1+\e}}\qquad \longrightarrow \qquad
A~\frac{1}{\e} ~\G(2\e) \label{2e2} \eeq The $1/\e^2$ term in
Fig.6b arises from \beq B~\frac{1}{\e}\int d^D k \frac{1}{(p+k)^2
(k^2)^{1+4\e}}\qquad \longrightarrow \qquad B~\frac{1}{\e}~
\G(5\e) \label{2e5} \eeq It is clear that setting $A+B=0$ is not
enough to cancel the  $1/\e^2$ poles.

 In order to check this general argument we
have computed the $1/\e^2$ divergent terms explicitly. At order
$g^{10}$  from the two-loop graphs shown in Fig.2, denoting with
$I_2$ the divergent integral in Fig.6a we have \beq -6(a_4+b_4)N^2
I_2\qquad \longrightarrow \qquad
-15~\z(5)~N^5~\frac{1}{(4\p)^6}(a_1-b_1)^4
~\frac{1}{(4\p)^4}\frac{1}{2\e^2} \label{2e2again} \eeq where we
have used the relation in (\ref{g8}).  In the same way from the
five-loop graphs shown in Fig.5, denoting with $I_5$  the
divergent integral in Fig.6b,  we obtain \beq 3(a_1-b_1)^4 N^5 I_5
\qquad \longrightarrow \qquad 3(a_1-b_1)^4 ~N^5
~\frac{1}{(4\p)^{10}}\frac{\z(5)}{\e^2} \label{2e5again} \eeq
Clearly the terms in  (\ref{2e2again}) and (\ref{2e5again}) do not
add up to zero and in fact they reproduce the mismatch anticipated
in (\ref{2e2}) and (\ref{2e5}) when $A+B=0$. Therefore at order
$g^{10}$ the cancelation of the $1/\e^2$ poles requires that (see
also (\ref{order1}) and (\ref{g8}))
 \beq a_1=b_1=1 \qquad \qquad
a_4+b_4=0 \label{e2poles} \eeq

Once the conditions in (\ref{e2poles}) have been imposed, at the
order $g^{10}$ all the $1/\e$ divergences from diagrams at five
and two loops are automatically cancelled. Thus at this order the
only divergence comes from the one-loop bubble and we are forced
to impose \beq a_5+b_5=0 \eeq

Before proceeding to the next 
order $g^{12}$, let us note that
this pattern of cancelling divergences between the one-loop bubble
in Fig.1a and the four-loop diagram in Fig.4 will repeat itself at
order $g^{16}$, while the cancelation of the  $1/\e^2$ poles will
show up  at the order $g^{18}$ and will involve again the diagrams
at two and five loops that we have just considered. Indeed at this
stage from the divergent contribution of  the four-loop diagrams,
using the conditions imposed so far on the coefficients of the
expansions in (\ref{expansion}),  the first divergence will be
proportional to 
\beq 
[(a_2-b_2)~g^4]^4=(2a_2)^4~g^{16} 
\label{gg16}
\eeq 
So for the time being, having ensured conformal invariance of the
theory up to the order $g^{10}$, we proceed and examine the
situation at six loops. The new divergent chiral diagrams are shown 
in Fig.7: they are all logarithmically divergent. 

\begin{figure}[ht]
\begin{center}
\epsfysize=2.5cm\epsfbox{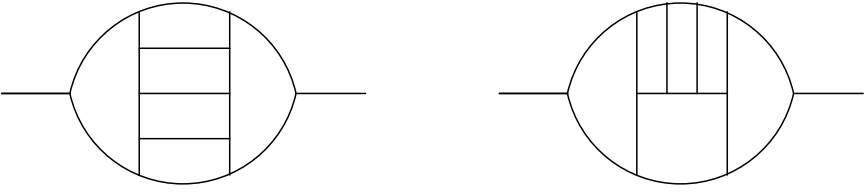}
\end{center}
\caption{New planar chiral diagrams at six loops}
\label{Fig7}
\end{figure} 

\noindent
Their color factor
is easily evaluated: it can be written in the following form \beq
(h_1^2+h_2^2)^6~+~(h_1^2-h_2^2)^4~(\frac{5}{3}h_1^4+\frac{2}{3}h_1^2
h_2^2+\frac{5}{3}h_2^4) \label{color6} \eeq Thus we find that in
the $g^2$  expansion the first nonvanishing term from the six-loop divergence
will be proportional to \beq [(a_2-b_2)~g^4]^4~ g^4=(2a_2)^4~
g^{20} \label{six} \eeq Thus once again to the order $g^{12}$ the
only divergence arises from the one-loop bubble and its
cancelation requires \beq a_6+b_6=0 \eeq We keep on going and look
for divergent terms at the order $g^{14}$. The diagrams at seven
loops have a color factor proportional to \beq
(h_1^2+h_2^2)^7~+~(h_1^2-h_2^2)^4~(3 h_1^6+ 5 h_1^4
h_2^2+ 5 h_1^2 h_2^4+ 3 h_2^6) \label{color7}
\eeq which, using the expansion in (\ref{expansion}), gives as
first relevant term \beq [(a_2-b_2)~g^4]^4 ~g^6=(2a_2)^4 ~g^{22}
\label{seven} \eeq Therefore once again the only divergence at the
order $g^{14}$ comes from one loop and leads to the condition \beq
a_7+b_7=0 \eeq

In accordance with the general discussion around equations (\ref{natural}, 
\ref{functions}) and what we have found by the explicit calculations
we have reported  up to seven loops, we write the $L$--loop
color structure of the pure chiral divergent diagrams in the
following form 
\beq
(h_1^2+h_2^2)^L~+~(h_1^2-h_2^2)^4~[\a(h_1^2)^{L-4}+\b(h_1^2)^{L-5}h_2^2+
\dots+\g(h_2^2)^{L-4}] 
\label{factoriz} 
\eeq
We note that the only arbitrary assumption with respect to the general form 
that
one can infer from (\ref{natural}, \ref{functions}) is the absence of a term
proportional to $(h_1^2-h_2^2)^2$. Even if we do not have a general 
argument for the absence of such a term we are very well supported by the 
results up to seven loops illustrated so far.   

If we take into account the conditions found
so far for the coefficients in (\ref{expansion}), then
(\ref{factoriz}) immediately implies that the various diagrams at
L loops will give contributions in the $g^2$ expansion whose first
relevant term is proportional to \beq [(a_2-b_2)~g^4]^4
~g^{2L-8}=(2a_2)^4 ~g^{2L+8} \label{Lloop} \eeq The conclusion is
that diagrams at six loops or higher will start contributing at
the earliest when we reach order $g^{20}$, as we have explicitly
seen in (\ref{six}) and (\ref{seven}). Therefore if we now turn
 to the order $g^{16}$,  as previously anticipated, the only divergent 
contributions come   from the one-loop bubble proportional
to $a_8+b_8$ and  from the four-loop diagram proportional to
$a_2^4$ (see eq. (\ref{gg16})). 
In order for the divergences to cancel at this order we
have to require \beq {\cal O}(g^{16}): \qquad \quad A_8=0 \quad
\longrightarrow \quad
a_8+b_8-\frac{5}{2}~\z(5)~N^3~\frac{1}{(4\p)^6}(a_2-b_2)^4=0
\label{g16} \eeq Going up to the order $g^{18}$ we have to cancel
the $1/\e^2$ poles from the two and five-loop diagrams: following
the same steps as before we are forced to impose 
\beq
a_8+b_8=0 \qquad \qquad a_2=b_2=0 
\label{a2} 
\eeq 
With these conditions on the
coefficients in the expansion (\ref{expansion}), at order $g^{18}$
the  $1/\e$ poles come only from the one-loop bubble and they
cancel out once \beq a_9+b_9=0 \eeq Since in (\ref{a2}) we have
imposed $a_2=0$, automatically we find that  the various
divergences from six, seven, $\dots,L$-loop diagrams are pushed up
\bea
&&{\rm 6~~loops}\qquad  \longrightarrow \qquad [(a_3-b_3) ~g^6]^4~g^4=
(2a_3)^4~ g^{28}\nonumber\\
&&{\rm 7~~loops} \qquad \longrightarrow \qquad [(a_3-b_3)~ g^6]^4~g^6=
(2a_3)^4 ~g^{30}\nonumber\\
&&~~~~~\dots\qquad\quad\dots\quad\qquad\dots\nonumber\\
&&{\rm L~~loops} \qquad \longrightarrow \qquad [(a_3-b_3)~
g^6]^4~g^{2L-8}=(2a_3)^4~ g^{2L+16} \eea
 It becomes clear that everything is ruled by the cancelation of divergences
at one  and four loop and by the subsequent cancelation of the
$1/\e^2$ poles at two and five loops. This happens at the order
$(g^2)^{4k}$ and at the order $(g^2)^{4k+1}$ respectively. 
The new chiral graphs at six loops and higher never enter the game due to the 
specific form of  their color structure as in (\ref{factoriz}).
The mechanism works as follows: up to the order $(g^2)^{4k-1}$ we find that
 the coefficients have to satisfy
 \beq a_1=b_1=1
\qquad\qquad a_{j-1}=0\qquad\qquad
a_{4j-1}+b_{4j-1}=0~~~~~j=2,\dots,k \eeq At ${\cal
O}((g^2)^{4k})$ in order to cancel the divergent contributions from one 
and four loops we have to impose 
\beq
a_{4k}+b_{4k}-\frac{5}{2}~\z(5)~N^3~\frac{1}{(4\p)^6}(a_k-b_k)^4=0
\label{g8k} \eeq 
Then at ${\cal O}((g^2)^{4k+1})$ the divergences from two and five loops 
need to be cancelled and we are forced to require 
\beq
a_{4k}+b_{4k} =0 \qquad\qquad a_k=b_k=0 \eeq 
Finally this leads to
 \beq a_1=b_1=1 \qquad\qquad
a_n=b_n=0~~~~~n=2,3,\dots
\label{finalcoeff}
 \eeq
These conclusions have been drawn based on the general expression given in 
(\ref{factoriz}) for the color structure of pure chiral diagrams where
we have assumed the absence of a term quadratic in $(h_1^2 - h_2^2)$. 
Now which control do we have on this assumption in the higher-loop 
divergent chiral diagrams? We have computed explicitly all the color 
structures up to ten loops; with the help of Mathematica we have evaluated 
the color factors of arbitrarily chosen higher-loop graphs; in addition we 
have explicit formulas for several classes of chiral diagrams. We have found  
consistently that all of them can be cast in the form given in 
(\ref{factoriz}).  

The conditions (\ref{finalcoeff}) on the coefficients tell us that the 
$\b$-deformed SYM theory is conformally invariant only for $\b$ real.

In the AdS/CFT dual description supergravity solutions associated to a 
complex parameter can be generated by completing the usual TsT transformation 
which leads to the Lunin--Maldacena background with S--duality transformations 
\cite{LM}. However, as discussed in \cite{FRT}, S--duality transformations 
might affect the 2d conformal invariance of the string sigma--model and this 
would require the appearance of $\a'/R^2$ corrections to the classical 
superstring action and then to the Lunin--Maldacena background. The fact that
a complex deformation parameter might be problematic is also signaled by
the appearance of singularities in the deformed metric when an imaginary part 
of $\b$ is turned on \cite{GN}. Therefore, the result we have obtained on
the field theory side seems to be in agreement with AdS/CFT expectations.

We stress that our investigation has been carried on perturbatively, ignoring
completly possible nonperturbative effects. In particular, we have assumed 
the gauge coupling constant to be real in order to avoid the 
presence of nontrivial 
instantonic effects \cite{instantons}. 
It would be interesting to extend our analysis to $g$
complex and  to understand if the embedding of all the couplings in a complex 
manifold leads to nontrivial superconformal conditions.  

\vspace{0.8cm}

\medskip

\section*{Acknowledgements}
\noindent This work has been supported in part by INFN, PRIN prot.2005024045-002
and the European Commission RTN program MRTN--CT--2004--005104.

\end{document}